\documentclass[twoside]{article}
\usepackage{graphicx}
\usepackage{fancyhdr}
\usepackage{multicol}
\usepackage{styl-ap}

\begin{document}
\title{Radioactivity and fragmentation of UHECR:\\ Correlating gamma anisotropy and neutrino PeV events }
\author{Daniele Fargion\work{1},\work{2} }
\workplace{Physics Department Rome University 1, Sapienza,
\next
INFN Sezione di Roma 1, Rome, Italy
}
\mainauthor{daniele.fargion@roma1.infn.it}
\maketitle
\begin{abstract}%
UHECR (Ultra High Energy Cosmic Rays) were expected to be protons, to fly straight and to suffer from a GZK (opacity on CMB radiation) cut off.
AUGER  suggested in 2007 that such early UHECR anisotropy was compatible with the foreseen Super-Galactic plane, while both HIRES and AUGER confirmed this \emph{apparent} GZK cut-off in the spectra. However, the same AUGER composition since 2007 had favored nuclei (and not nucleons). The recent absence of narrow angle clustering in UHECR maps, as it should be expected for protons, the missing of events along the nearest Cluster Virgo, the wide spread (nearly $16^{o}$ angle) of UHECR along Cen A are in disagreement with the first proton-UHECR AUGER understanding. Since 2008 we had claimed a light nuclei role for the Cen A crowded area. However the ICECUBE absence of TeV neutrino clustering or anisotropy, its spectra steepening favors mostly a ruling  atmospheric neutrino noise up to tens TeV. However  two recent PeV neutrino event,  cannot easily coexist or been extrapolated with such an atmospheric ruling scenario, or with GZK (either nucleons or nuclei) secondary expected  spectra. Finally tens of TeV gamma anisotropy in ARGO-MILAGRO-ICECUBE maps  may hardly be associated with known hadronic sources. We imagine such anisotropy ruled by diffused gamma secondaries, shining along UHECR bending and flight: radioactive light and heavy UHECR nuclei, while scattering or  decaying in flight, may paint in the sky (by gamma, electrons and neutrinos secondaries) their trajectories and bending, connecting UHECR spread events  with $\gamma$ TeV anisotropy, and also offering a very realistic source of first, otherwise puzzling, observed PeV neutrinos.

\end{abstract}


\begin{multicols}{2}
\section{Introduction}
Ultra High Energy cosmic Rays (UHECR) and UHE neutrino Astronomy are  two of the hot frontiers of present High Energy Astrophysics.
For half a century we have expected to detect UHECR  from the near Universe (because of the photon-nucleon GZK cut off opacity due to cosmic  radiation) and along their straight trajectory, because of the highest proton rigidity. Almost five years ago, the AUGER experiment claimed in its abstract that \emph{" they demonstrated a correlation between the arrival directions of cosmic rays with energy above 60 EeV electron volts and the positions of active galactic nuclei (AGN) lying within 75 mega-parsecs"}. Therefore \emph{" They rejected the hypothesis of an isotropic distribution of these cosmic rays with at least a $99\%$ confidence level from a prescribed a priori test. The correlation  observed is compatible with the hypothesis that the highest-energy particles originate from nearby extragalactic sources whose flux has not been substantially reduced by interaction with the cosmic background radiation (GZK cut off). AGN or objects having a similar spatial distribution are possible sources".}  However more recent maps seem to dismiss  this possibility. UHECRs are not well correlated to the Super-Galactic Plane; they do not cluster where nearest AGN are located. Following AGASA isotropy maps many attempts were made in the past(eg.the  Z-Showering model \cite{Fargion1997} or heavy relic annihilation \cite{Dubrovich}) to solve the puzzle; moreover  AUGER now claim that UHECR are most probably nuclei; we have suggested that UHECR are the lightest nuclei  for Cen A and/or the heaviest ones for most other events; we also explain here that we believe they are mostly galactic and partially within the inner Local Group. Indeed, very recent UHECR maps and their possible secondaries multiplet  offer a relevant Cen A role as the main unique extragalactic UHECR source; we suspect that that several minor clustering are suggesting few local galactic sources. At the highest edges (EeV) neutrinos may trace the same proton  GZK traces, if UHECR (as most still believe) are protons; in such proton-UHECR model, cosmogenic PeVs neutrinos will be almost undetectable today (contrary to recent 28 Icecube $\nu$ events) ; neutrino will be even poorer if they derive from photo-nuclear fragments of GZK nuclei.
 On the contrary, as we shall show, PeV neutrinos may be born from UHECR EeV radioactive decay nuclei or from UHECR EeV nuclei  scattering on nucleons in dense galactic clouds. Icecube and AUGER until now failed  to find the EeV GZK neutrino (by proton photo-pion decay) tracing UHECR clustering source. However recent two isolated recent PeV (with additional 26 tens-hundred TeV ) events in ICECUBE are probably not atmospheric neutrinos \cite{Fargion13}  also no EeV GZK neutrinos;  UHECR made by radioactive nuclei whose relativistic decay shine at PeV-TeVs or from UHECR EeV nuclei  scattering on nucleons at hundred TeV pions may fit the picture. We imagine the UHECR source as the GRB-SN Jet sources spraying nuclei.
  UHECR in the latest non relativistic stages may also be traced by eventual radioactive decay (Co, Ni) at MeV; however in the earlier inner jetted  boosted stage the UHECR tracks may be source of correlated TeVs gamma anisotropy as well  be source of PeV-TeVs neutrinos recently discovered. UHECR may therefore trace  UHE TeV gamma maps, while  ICECUBE showering PeV neutrino relics, probably of tau and electrons flavor, may be their associated signature.
\subsection{A century of cosmic rays}
The last two decades of high energy astrophysics have puzzled us with  Ultra High Energy Cosmic ray (UHECR) spectra and mysterious maps. After the first huge Fly's Eye event (1991), the later  absence of GZK cut off in AGASA UHECR made theoreticians speculate about a new physics. The more recent HIRES (2005) and AUGER data seemed to confirm GZK cut off spectra, and since 2007 (AUGER,2007) also the anisotropy map was ,apparently, pointing to nearby GZK nearby sources. However, more recent maps and more composition signatures are in strident contradiction with this goal: today UHECR by AUGER does not much follow the UHECR nearby (GZK cosmic volume) mass distribution. However some remarkable ($20\%$) event clustering remains along nearest AGN: Cen A. However, the common proton currier UHECR (single charge) would be more collimated $3^{o}-4^{o}$, while the clustering is spread by $10^{o}-15^{o}$ along the Cen A source. The spread is enlonged vertically with respect to the galactic plane.
We suggested that such tail events are made mostly of UHECR He nuclei. We argued four years ago that their fragments in flight,  would fragment into fifteen EeV lower energy secondaries (protons,neutrons,D); such a train of UHECR multiplets was (probably) observed on 2011 by AUGER along  with Cen A, with low ($3 \cdot 10^{-5}$ ) probability occurring by chance. More recent clustering of UHECR along different radio and gamma maps forced us to imagine a way to associate UHECR with TeV gamma anisotropy. We found that this is possible in the assumption that UHECR are partially  the lightest nuclei (extragalactic) and mostly in a very local universe, or even galactic. This UHECR connection with TeV gamma spread maps is possible because lightest UHECR secondaries from Cen A are  also formed by neutrons whose decay in flight showers into tens PeV electrons followed by PeV-TeV gamma; the heaviest UHECR nuclei in our galaxy may be radioactive or scattering on gas and they may also decay or pion producing in flight in gamma and electrons radiated by Lorentz boost into PeVs energy ranges. UHECR may contain both  (lightest and heaviest) nuclei offering this behavior. As a consequence, contrary to the generally accepted belief, we probably, have not yet observed the real GZK cut off.
\section{Main  lesson of UHECR }
There are some clear lessons from the most recent UHECR maps and the spectra and their composition:
1) UHECR are mostly not protons; indeed  there is not narrow multiplet   clustering within an angle of few degree  as expected for protons;
we found a doublet near Cen A and a twin highest hundred EeV along the Galactic Plane, but most few clustering are spread in a long smeared tail. The AUGER composition shows more nuclei than just protons. Hires, TA are compatible with protons and/or light nuclei.
2) Virgo, the nearest and most abundant crowded region of AGN sources within our GZK volume is absent,  see Fig. \ref{Fargion-fig2};
  (this region of the sky is partially suppressed by AUGER detection, but the paucity of events paucity raises a question):
light nuclei as a UHECR currier, being nearly opaque to CMB radiation from distances Virgo, may explain the absence of a major UHECR component.
3) The size (and the elonged strip) form of some UHECR clustering  favors a light-heavy charge (heavy nuclei) bent by galactic magnetic fields,  see Fig. \ref{Fargion-fig1}, see Fig.\ref{Fargion-fig3}.
4) The absence of a Super-galactic plane correlation may favor a more local sky, as a few tens of EeV multiplets along Cen A and the Magellanic Stream suggest. Indeed the Planck dust cloud region, mostly galactic, is where UHECR occur abundantly,  see Fig. \ref{Fargion-fig3}.
  There is also a remarkable correlation between UHECR events and the neutral H map, see Fig. \ref{Fargion-fig4}.
5) The observed UHECR GZK cut off may  very well be an UHECR composition change steepening at different energy ranges. The expected GZK cut off may therefore be just an illusion.

   The highest PeV-TeVs neutrinos are not   probably of atmospheric nature \cite{Fargion13} and they are probably also not of resonant W boson nature. This neutrino paradox (muon paucity and GZK absence) may nevertheless ensure that
   UHE neutrino Astronomy is still born.

    However  we offer here additional ways to produce PeV neutrinos: one is at their birth along the source itself by hadronic secondaries \cite{Roulet-12}. This point-like source must be also observed by a growing inhomogeneity in neutrino maps. The very recently produced map (hundreds of thousands of events) in ICECUBE does not exhibit such clustering \cite{aguilar}. A different source of UHE neutrinos may arise during the SNR or GRB jet emission:
     UHECR heavy radioactive nuclei  may rule (as observed) the SN peak optical luminosity for days and weeks and  fast parental decay boosted at a high Lorentz factor (a millions-billion) of UHECR nuclei like Ni,
     may also shine for thousands of years in  PeV neutrino and electron tails, as discussed in the next section, at lower TeVs  gamma energies
      because of  PeV electron synchrotron radiation. Also UHECR nuclei may hit gas along their flight spreading secondaries in wide sky regions.  These radioactive tails, like the tail observed by ARGO in TeVs along the Crab
      and in  nearby areas, maybe smeared and non point-like, explaining both the presence of PeV neutrino events and the absence
      of the tens TeV neutrino clustering. TeV neutrino anisotropy might rise smeared as for the ARGO-MILAGRO anisotropy at a flux deviation as small   as $10^{-4}$, corresponding to a flux density of the order of $\simeq 30-50$ eV $cm^{-2} s^{-1}sr^{-1}$ comparable to recent observed one \cite{Fargion13}.
    The last  lessons from AUGER are the low multiplet clustering in tens of EeV energy range, meaning and fate of which is simply ignored by most authors. Two of these multiplet clusters along Cen A are shown in Fig. \ref{Fargion-fig1}.
    \subsection{Bending for He UHECR and fragments at 20 EeV along Cen A}
 The very  recent multiplet clustering published on 2011 by AUGER at twenty EeV  contains just three apparently isolated trains of events  with  (for the AUGER collaboration) no statistical meaning. \cite{Auger11}. Indeed, they apparently  point to unknown sources (See Fig. \ref{Fargion-fig1}). However, the crowding of the two train multiplet tail centers inside a very narrow disk area focused about the rarest Cen A UHECR source is remarkable, \cite{Fargion2011}.  If UHECR are  made by protons (as some AUGER and TA author believe), they will not naturally explain such a tail structure because these events do not cluster more than a few degrees, unlike the observed UHECR and the associated multiplet. Our He-like UHECR fit the AUGER as well as the HIRES and TA composition traces. The He secondaries  split in half (or a fourth) energy fragments along the Cen A tail  the presence of which has  being foreseen, published (and ignored) many times  in recent years\cite{Fargion2009-2010},\cite{Fargion2011}. Indeed, the  dotted circle around Cen A containing two (of three) multiplets  has a radius as small as $7.5^{o}$ extending in an area that is as small as  $200$ square degrees, below or near $1\% $ of the  AUGER observation sky. The probability that two out of three sources fall inside this small area is offered by the binomial distribution. $$ P (3,2) = \frac{3!}{2!} \cdot (10^{-2})^{2} \cdot \frac{99}{100}\simeq 3 \cdot 10^{-4}.$$ Moreover the same twin tail of the events is aligned almost exactly $\pm 0.1 $ rad along the UHECR train of events toward Cen A. Therefore the UHECR  multiplet alignment  at twenty EeV has an a priori  probability as low as $ P (3,2) \simeq 3 \cdot 10^{-5}$ of following the foreseen signature\cite{Fargion2011}.
  The incoherent random angle bending (2) along the galactic plane and arms, $\delta_{rm} $, while crossing along the whole Galactic disk $ L\simeq{20\cdot kpc}$  in different (alternating) spiral arm fields   and within a characteristic  coherent length  $ l_c \simeq{2\cdot kpc}$ for He nuclei is $\delta_{rm-He}\simeq $ $$\simeq{16^\circ}\cdot \frac{Z}{Z_{He^2}} \cdot (\frac{6\cdot10^{19}eV}{E_{CR}})(\frac{B}{3\cdot \mu G})\sqrt{\frac{L}{20 kpc}} \sqrt{\frac{l_c}{2 kpc}}$$ The heavier  (but still light) nuclei bounded from Virgo might also be Li and Be: $  \delta_{rm-Be} \simeq $
  $$\simeq{32^\circ}\cdot \frac{Z}{Z_{Be^4}} \cdot (\frac{6\cdot10^{19}eV}{E_{CR}})(\frac{B}{3\cdot \mu G})\sqrt{\frac{L}{20 kpc}}
\sqrt{\frac{l_c}{2 kpc}}.$$  It should be noted that the present anisotropy above GZK \cite{Greisen:1966jv} energy $5.5 \cdot 10^{19} eV$ (if extragalactic)  might leave a tail of signals: indeed the photo disruption of He into deuterium, tritium, $He^3$ and protons (and unstable neutrons), arising as clustered events at a half or a quarter (for the last most stable proton fragment) of the energy:\emph{ protons being with a quarter of the energy but a half of the charge of the He parent may form a tail  smeared around Cen-A at a twice larger angle} \cite{Fargion2011}. We suggested looking for correlated tails of events, possibly in  strings at low $\simeq 1.5-3 \cdot 10^{19} eV$ along the Cen A train of events. \emph{It should be noted that Deuterium fragments have one half of the energy and mass of Helium: Therefore D and He spot are bent in the same way and overlap into UHECR circle clusters} \cite{Fargion2011}.  Deuterium is  even more bounded in a very local Universe, because of its fragility (explaining the absence of  Virgo). In conclusion, He like UHECR may be bent by a characteristic angle as large as $\delta_{rm-He}  \simeq 16^\circ$; its expected lower energy Deuterium or proton fragments at half energy ($30-25 EeV$) are also deflected accordingly at ($\delta_{rm-p}  \simeq 16^\circ$); the last traces of protons at a quarter of the UHECR energy, around twenty EeV energy, will be  bent and spread within ($\delta_{rm-p}  \simeq 32^\circ$), exactly within the observed Cen A UHECR multiplet, see Fig. \ref{Fargion-fig1}.\\

\begin{myfigure}
\centerline{\resizebox{70mm}{!}{\includegraphics{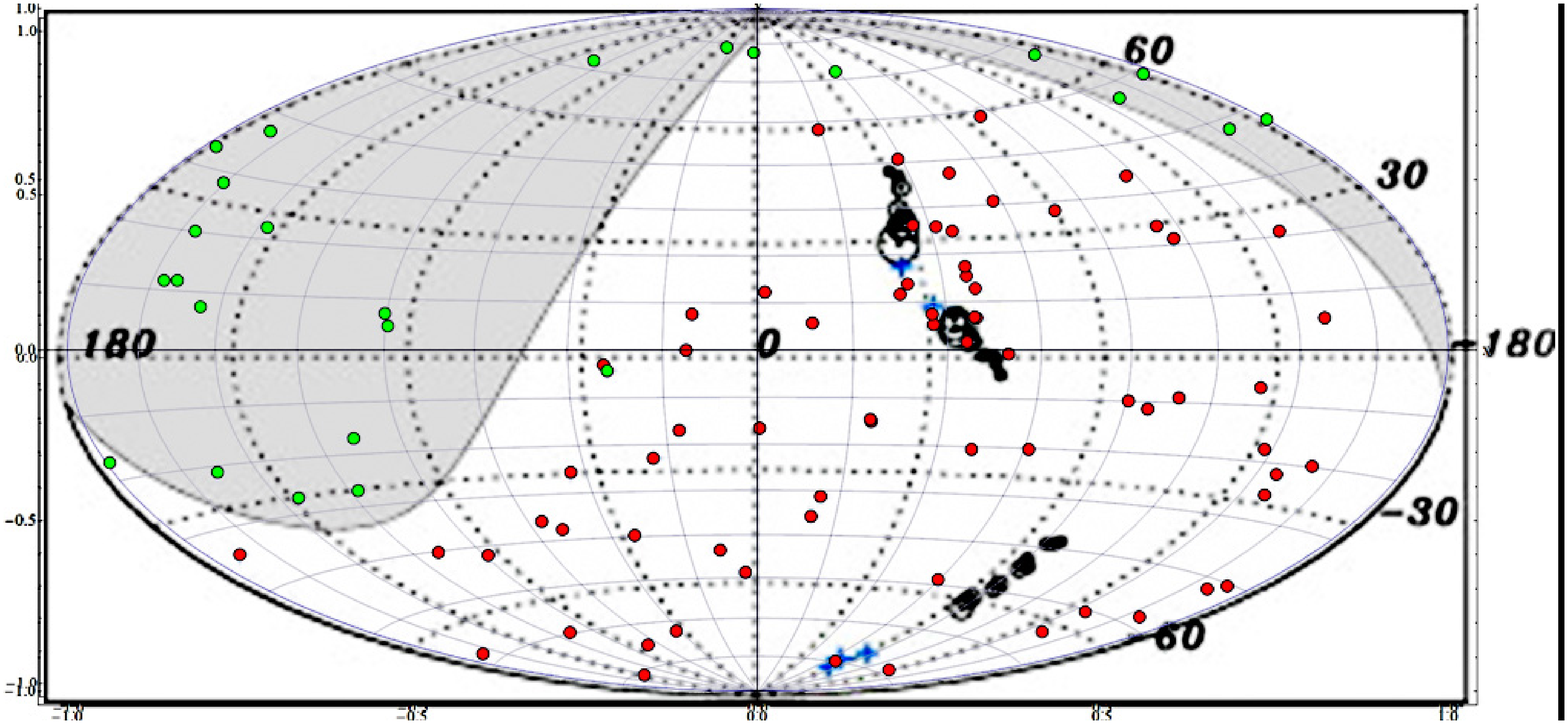}}}
\caption{The very last UHECR event maps for both AUGER (red disk) and TA (Telescope Array)(green disk)  events. AUGER is in the south equatorial side and TA is in the North side. Three multiplet at 20 EeV are also shown.
Note the crowding along the unique nearest AGN Cen A; note also possible multiplets crowding along the Large and Small Magellanic Clouds.}
\label{Fargion-fig1}
\end{myfigure}

\begin{myfigure}
\centerline{\resizebox{70mm}{!}{\includegraphics{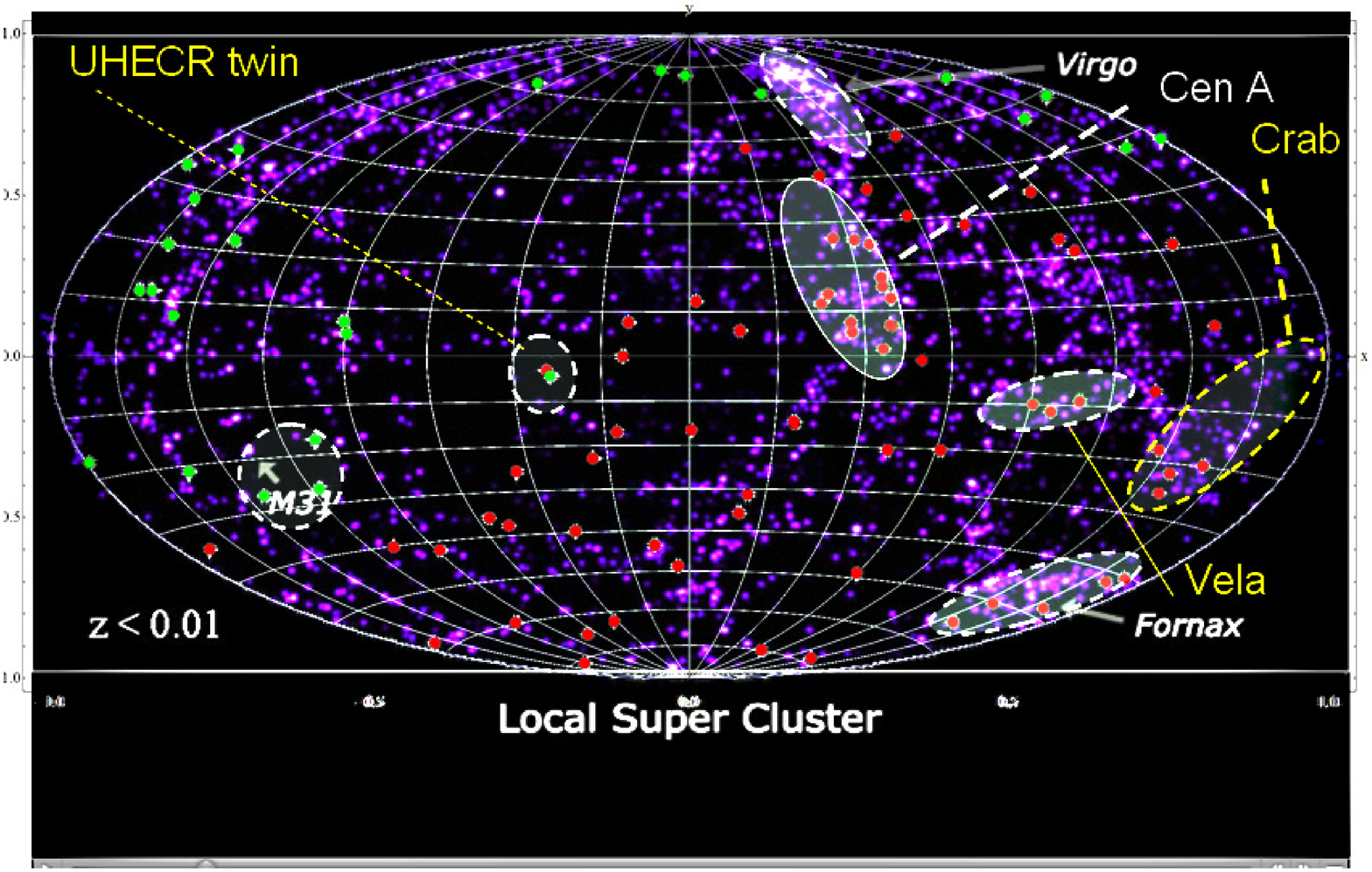}}}
\caption{The infrared cosmic map with the UHECR events (AUGER and TA); note the remarkable absence of UHECR at the nearest galaxy cluster, Virgo;
the presence of a quarter of AUGER events within a narrow sky area along Cen A; the possible crowding along the nearest Fornax  dwarf galaxy source; along the Orion and Crab areas, and a rare triplet toward M31 and a triplet near Vela, our brightest Fermi gamma SNR and gamma PSR. }
\label{Fargion-fig2}
\end{myfigure}

\begin{myfigure}
\centerline{\resizebox{70mm}{!}{\includegraphics{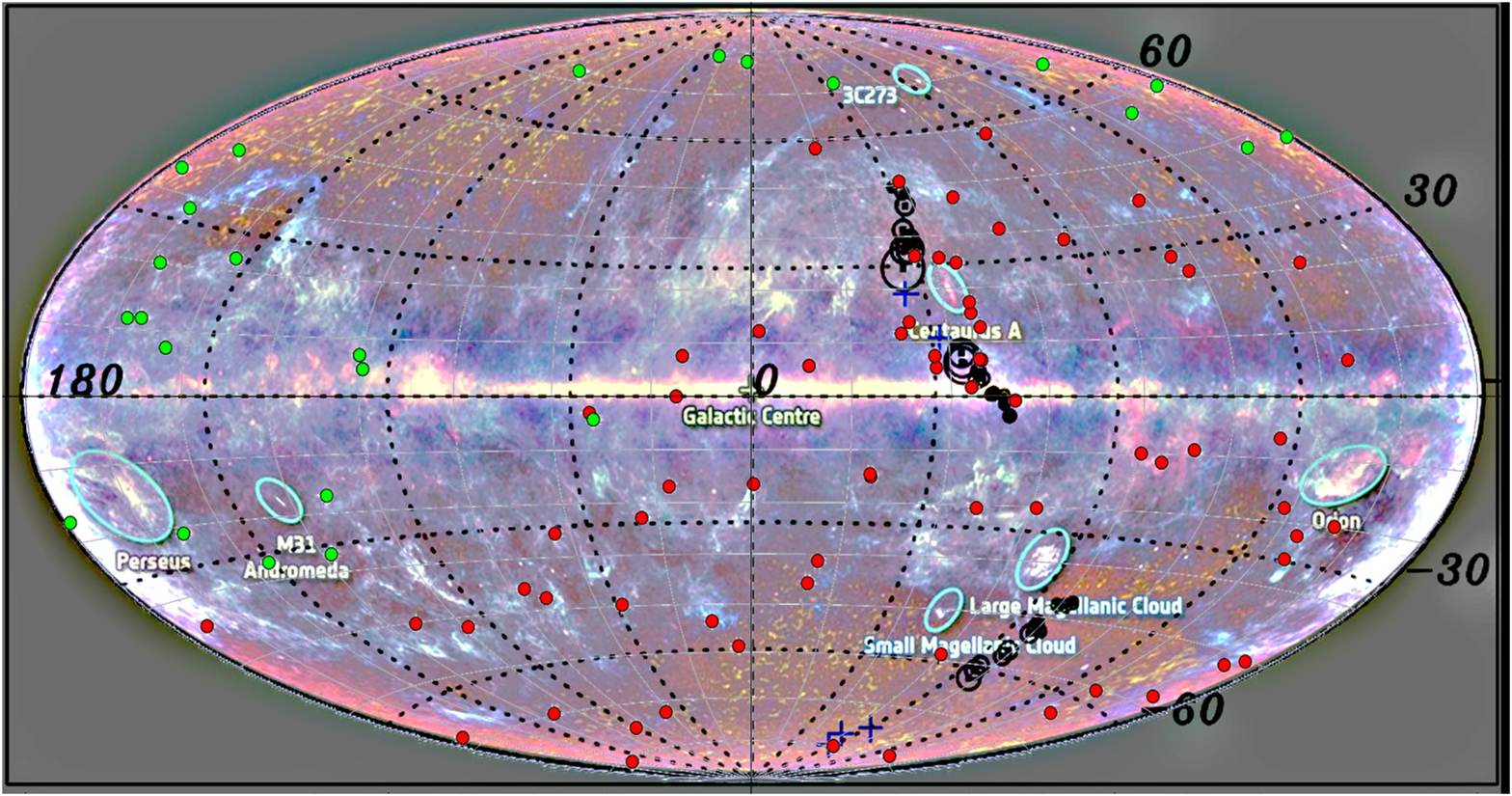}}}
\caption{The Planck infrared last maps with both AUGER and TA events. The paucity of UHECR events inside the infrared map, with little or none galactic dust presence, is remarkable : this suggest the main role of galactic sources and the gas target role in scattering and fragmenting UHECR}
\label{Fargion-fig3}
\end{myfigure}

\begin{myfigure}
\centerline{\resizebox{70mm}{!}{\includegraphics{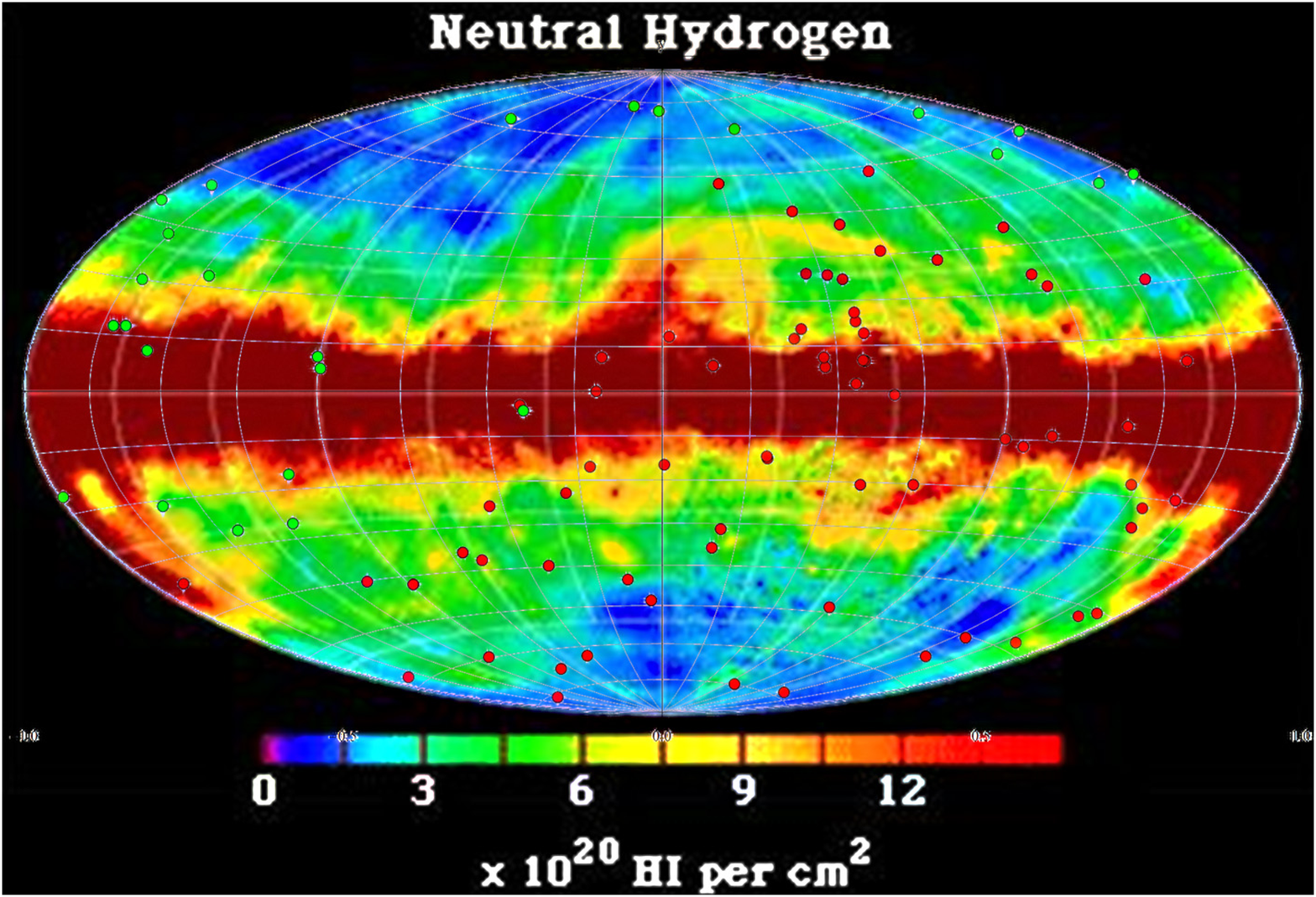}}}
\caption{The neutral hydrogen map with the UHECR events: as in the previous Planck map, the UHECR are almost absent where there is no galactic hydrogen gas. This peculiarity may favor a very galactic UHECR component.}
\label{Fargion-fig4}
\end{myfigure}

\section{TeV Gamma and UHECR }
       In recent  UHECR maps we have noted first hint of a galactic source   arising as a UHECR triplet \cite{Fargion09b}. The hint of the $Al^{26}$ gamma map traced by Comptel somehow overlapping with UHECR events at 1-3 MeV favors a role of UHECR radioactive elements (as $Al^{26}$).   The most prompt radioactive nuclei are the  $Ni^{56}$, $Ni^{57}$ (and $Co^{56}$, $Co^{60}$ ), made by Supernova (and possibly by their collimated GRB micro-jet components, ejecta in our own galaxy). Similar radioactive traces may arise by UHECR scattering on dense gas clouds.  Indeed in all SN Ia models, the decay chain  $Ni^{56}\rightarrow Co^{56} \rightarrow Fe^{56}$ provides the primary source of energy that powers the supernova optical display even days and weeks later the explosion. $Ni^{56}$ decays by electron capture and the daughter $Co^{56}$  emits gamma rays by the nuclear de-excitation
process; the two characteristic gamma lines are at $E_{\gamma} =$ 158 keV and $E_{\gamma}=$ 812 keV  rspectively. Their half lifetimes are spread from $35.6$ h for  $Ni^{57}$ and $6.07$ d. for   $Ni^{56}$. However there are also more unstable radioactive rates, as  for $Ni^{55}$ nuclei whose half life is just $0.212$ s or $Ni^{67}$, whose decay is $21$ s. Therefore we may have an apparent boosted  UHECR ($\Gamma_{Ni^{56}} \simeq 10^{9}$) lifetime spread from $2.12 \cdot 10^{8}$ s or $6.7$ years (for $Ni^{55}$) up to nearly  $670$ years (for $Ni^{67}$) or $4$ million years for  $Ni^{57}$. EeV and PeV radioactive UHECR or their fragment may also play role in gamma and neutrino emission. This consequent wide range of lifetimes guarantees a long life activity on the UHECR radioactive traces. The arrival tracks of these UHECR radioactive heavy nuclei may be  widely bent, as shown below, by galactic magnetic fields. Among the excited nuclei to mention  for the UHECR-TeV connection is $Co_{m}^{60}$  whose half life is $10.1$ min and whose decay gamma line is at $59$ keV. At a boosted nominal Lorentz factor $\Gamma_{Co^{60}}= 10^{9}$, we obtain $E_{\gamma}\simeq 59 $ TeV ; note that a gamma air-shower exhibits a smaller secondary muon abundance with respect to the equivalent hadronic abundance; therefore a gamma  simulates a ($10\%$) hadronic shower ( $E_{gamma-hadron}\simeq 6 $ TeV)  corresponding closely to the observed ICECUBE-ARGO anisotropy \cite{ARGO}. The decay boosted lifetime is $19000$ years, corresponding to  6 kpc distance. Therefore $Co_{m}^{60}$ energy decay traces, lifetime and spectra fit well within  the present UHECR-TeV connection for nearby galactic sources as Vela and (probably) Crab. Other radioactive scattering trace, usually at lower energy may also shine at hundreds or tens of TeV or below by inverse Compton and synchrotron radiation. Therefore their UHECR bent parental nuclei may  also shine in TeV Cosmic ray signals. In beta decay processes, electrons and neutrinos are also born, providing a new diffused gamma and PeV neutrino source.

\section{ UHECR galactic bending  for $Ni^{57}$ }
Cosmic Rays are blurred by magnetic fields. UHECR also suffer from Lorentz force deviation. This smearing maybe a source of UHECR features, mostly along Cen A. There are at least three mechanisms for magnetic deflection along the galactic plane, a sort of galactic spectroscopy of UHECR \cite{Fargion2008}. Magnetic  bending by extra-galactic fields is in general negligible in comparison with galactic bending.  Late nearby (almost local) bending by a nearest coherent galactic arm field, and random bending by turbulence and  random deflection along the whole plane inside different arms:\\
(1) the coherent Lorentz angle bending $\delta_{Coh} $ of a proton (or nuclei) UHECR (above GZK \cite{Greisen:1966jv}) within a galactic magnetic field  in a final nearby coherent length  of $l_c = 1\cdot kpc$ is: \\$ \delta_{Coh-p} .\simeq{2.3^\circ}\cdot \frac{Z}{Z_{H}} \cdot (\frac{6\cdot10^{19}eV}{E_{CR}})(\frac{B}{3\cdot \mu G}){\frac{l_c}{kpc}}$\\
(2) the random bending by random turbulent magnetic fields, whose coherent sizes (tens of parsecs) are short and whose final deflection angle is  smaller than others are ignored here;\\
(3) the ordered multiple UHECR bending along the galactic plane across and by alternate arm magnetic field directions whose final random deflection angle is remarkable and discussed below.\\The bending angle value is  quite different for a heavy nucleus such as a UHECR from Vela whose distance is only $0.29$ kpc:
 $\delta_{Coh-Ni} \simeq {18.7^\circ}\cdot \frac{Z}{Z_{Ni^{28}}} \cdot (\frac{6\cdot10^{19}eV}{E_{CR}})(\frac{B}{3\cdot \mu G})({\frac{l_c}{0.29 kpc})}$\\
  Note that this spread is  able to explain the nearby Vela TeV anisotropy (because of the radioactive emission in flight) area around its correlated UHECR triplet.  There is an extreme possibility: that a Crab pulsar at a few kpc is feeding the TeV anisotropy connecting with a gate its centered disk to a wider extended region where some UHECR are clustering, see last Fig.\ref{fargion-fig5}. From far Crab distances the galactic bending is:\\  $\delta_{CohNi} \simeq {129^\circ}\cdot \frac{Z}{Z_{Ni^{28}}} \cdot (\frac{6\cdot10^{19}eV}{E_{CR}})(\frac{B}{3\cdot \mu G})({\frac{l_c}{2 kpc})}$\\
  Note that such a spread is able to explain the localized TeV anisotropy born in Crab (2 kpc) apparently  extending  around an area near Orion, where  spread UHECR events also seem to be clustered,  see last Fig.\ref{fargion-fig5}. Such heavy iron-like  (Ni,Co) UHECR , because of the big charge and large angle bending, are mostly bounded inside a Galaxy, as well as in a Virgo cluster, possibly explaining the  absence of UHECR in that direction, see Fig.\ref{Fargion-fig2}. The possible galactic component of UHECR is suggested by the correlated dark Hydrogen and dust map with the UHECR distribution:  see Fig.\ref{Fargion-fig3}, Fig.\ref{Fargion-fig4}.

\begin{myfigure}
\centerline{\resizebox{90mm}{!}{\includegraphics{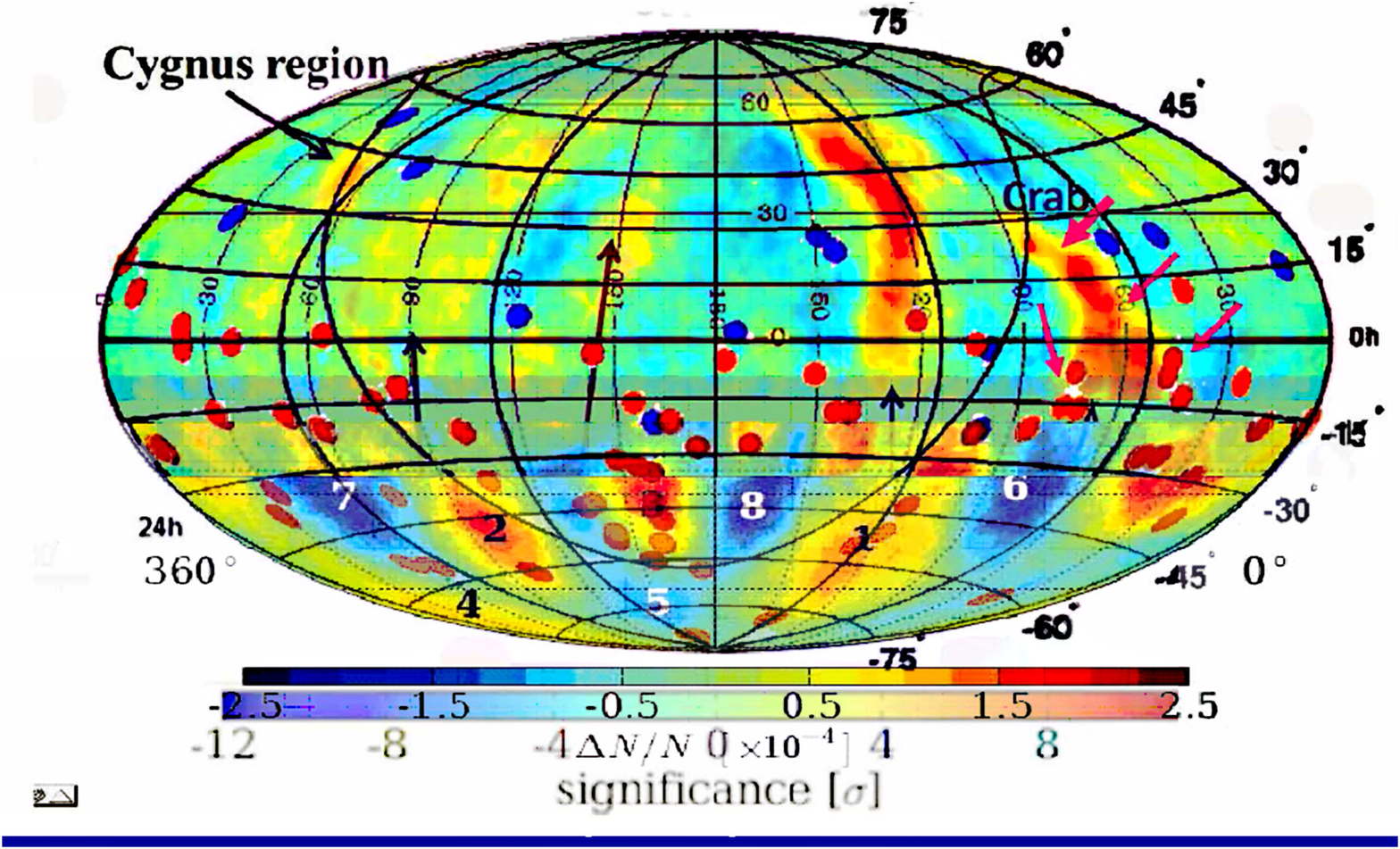}}}
\caption{The overlap of the 69 UHECR events by AUGER 2010 and 13 UHECR events by Hires in celestial coordinates.
 The clustering of UHECR along the red (dense) TeVs sky is remarkable: note the crowding of  events near area n.6, where Vela is located; note also the crowding of events to the left of  area n.8, where Cen A is located; note the crowding of events along the right side of the map, where the Fornax Dwarf Galaxy is located; note some clustering with the galactic plane and near the Crab and Orion regions; we foresee that  present and future TA and AUGER events will populate  the north and south sky area, respectively mostly along the TeVs anisotropy.}
\label{fargion-fig5}
\end{myfigure}

\section{Discussion}
The UHECR puzzle maybe  a cornerstone: the UHECR-Multiplet along Cen A, the absence of Virgo, the hint of a correlation with  Vela and  with galactic TeV anisotropy \cite{Desiati} \cite{ARGO}, might be in part solved by an extragalactic lightest nucleous, mainly He, from Cen A and in part by a galactic heavy radioactive UHECR nucleous or scattering fragment of such UHCR on gas. Our predicted \cite{Fargion09b} and observed \cite{Auger11} multiplet clustering by fragments (D,p) at half UHECR edge energy aligned   with Cen A favor such a reading of the UHECR map: He like UHECR  may be bent by a characteristic angle as large as  $\delta_{rm-He}  \simeq 16^\circ$ \cite{Fargion2011}, while their fragments multiplet  follow along a tail spread by a wider angle $\delta_{rm-p}  \simeq 32^\circ$ \cite{Fargion09b},\cite{Auger11}; also neutron beta decay from Cen A may feed a TeV correlated anisotropy.  Other UHECR spread events may  be due to a  dominant heavy radioactive nuclei component  $Ni^{56}$, $Ni^{57}$ and $Co^{56}$, $Co^{60}$,originating from galactic sources (old SNR-GRB relics, to day Soft Gamma Repeaters, SGRs), as also suggested  by relic $Al$ nuclei at rest in the gamma map \cite{Fargion2011}. UHECR Ni, Co maybe deflected by $18.7^{o}$ for Vela,  $128^{o}$ (or less) for Crab tuning within TeV inhomogeneities, made by boosted hundred keV gamma and beta positron decay, shining at TeVs. Also scattering of  EeV UHECR on gas proton may shower pions and photons at PeV-TeV energy. Inner galactic core UHECR are widely spread and hidden by magnetic fields in dense magnetic galactic core arms. However more clustering around ($\geq 20^{o}$) the galactic plane far from the core, is expected in future data. Magellanic cloud and stream may rise in UHECR maps. UHECR should rise around Cas A and Cygnus, observable by TA, in the North sky. A recent unique doublet at highest AUGER and TA energy toward  Aq X1 may be a new galactic source \cite{Troitsky}. The UHECR spectra cut off maybe not be due to the expected extragalactic GZK feature but to the more modest imprint of a galactic confinement and/or nuclei spectrography.    The UHECR radioactive beta decay in flight may trace in new $\nu_{\tau}$, neutrino astronomy or anisotropy, noise free,  related to astronomical (parasite oscillated) tau neutrino;   boosted tau (\emph{mini-double bang} \cite{Learned}, within a 5-10 meter size) in the Deep Core  or PINGU  may reveal high energy tau decay and shower ( similar to $10^{14}-10^{15}$ eV ones observable in ICECUBE \cite{Learned} with tiny elliptical anisotropic shape of their cascade event). Tau airshowers may also arise in Cherenkov beamed air-showers. \cite{Fargion1999}, \cite{FarTau}, as being searched in the ASHRA experiment, \cite{Aita11} or in fluorescence telescopes for higher tau energies \cite{FarTau},\cite{Feng02},\cite{Auger07}, \cite{Auger08}.
    \section{Conclusions}The discovery of TeVs-PeVs expected  Neutrino astronomy  may shed additional light on the nature, origination  and mass composition of UHECR, while  opening our eyes to their mysterious  sources. Tau neutrino astronomy by upward airshowers  and also additional ICECUBE showering at PeV with no muon (most probably due to an electron or tau cascade) \cite{aguilar}  \cite{Fargion13} may also offer first windows for the extraterrestrial neutrino traces, opening the road to the desired High Energy Neutrino Astronomy.
    Their connection to UHECR radioactive decay and scattering in flight is a possibility to be tested in future Cosmic ray anisotropy and UHECR maps. \\

\section{Dedication}
This article is devoted to the memory of our beloved bright nephew Daniela Aielet Nur Di Gioacchino Fargion,  born on 13rd December 1975, departed on  24th July 2012.

\end{multicols}
\end{document}